\documentstyle[prb,multicol,aps,epsf,psfig]{revtex}
\renewcommand{\narrowtext}{\begin{multicols}{2} \global\columnwidth20.5pc}
  \renewcommand{\widetext}{\end{multicols} \global\columnwidth42.5pc}

\begin{document}
\draft
\title{ Effects of Entanglement in Controlled Dephasing }

\author{Alessandro Silva and Shimon Levit}
\address{Department of Condensed Matter Physics, The Weizmann 
Institute of Science, 76100 Rehovot, Israel}

\maketitle
\begin{abstract}
{In controlled dephasing as a result of the interaction of 
 a controlled environment (dephasor) and the system under observation
(dephasee) the states of the two subsystems are entangled.
Using as an example the ``Which Path Detector'',
we discuss how the entanglement influences the controlled dephasing.
In particular, we calculate the suppression $\nu$ 
of A-B oscillations as a function of the 
bias $eV$ applied to the QPC and the coupling $\Gamma$ of the QD to the leads.
At low temperatures the entanglement produces 
a smooth crossover from $\nu \propto (eV/\Gamma)^2$, when $eV \ll \Gamma$ to 
$\nu \propto eV/\Gamma$, for $eV \gg \Gamma$.}
\end{abstract}

\vspace{0.5cm}
  
\narrowtext

\section{Introduction and main results}
\label{intr}

Recently, the progress in nanofabrication opened 
technological possibilities to fabricate mesoscopic devices consisting of pairs 
of {\it capacitively coupled} (no particle exchange) mesoscopic systems.
The primary motivation was to carry out controlled  experimental studies 
of dephasing in quantum electronic systems. In such experiments one system
plays the role of the "dephasor" which causes decoherence in the other system - 
the "dephasee". The possibility to experimentally adjust the type of dephasor 
and to vary its parameters should permit the production of
various patterns of decoherence in the dephasee.

The direct observation, by means of interference experiments, of
coherent transport through mesoscopic structures\cite{Yacobi}, is making it possible
to study these patterns. The first realization of controlled dephasing 
was the so called ``Which Path Detector'' (WPD)
~\cite{Bucks,Gurwitz,Aleiner}. It consisted
of an Aharonov-Bohm interferometer in a four terminal
configuration~\cite{Yacobi}  with a pinched  
Quantum Dot (QD) inserted in one arm playing the role of the dephasee. 
A biased Quantum Point Contact (QPC) located close to the QD 
played the role of the dephasor. The resulting suppression
of interference was measured by looking at the reduction of the 
amplitude of A-B oscillations.

It is useful to think of the dephasor-dephasee interaction as of a scattering
process. The source of dephasing (e.g., the suppression of the A-B oscillations
in  the WPD experiment) is the presence of {\it inelastic channels}, i.e., the
 possibility
to change the state of the dephasor~\cite{Imry}. 
It should be stressed that in this context 
``inelastic'' is not necessarily associated with ``energy transfer'', since in 
some situations it is possible
to change the state of the dephasor without energy exchange. 

In order to detect dephasing, one can ideally insert the dephasee  
in one arm of an interferometer~\cite{Imry} as in the WPD.
The wave function of the interferometer-dephasor system is 
$\mid \Psi \rangle=\mid \Psi_{ref} \rangle+\mid \Psi_{d} \rangle$, 
the two terms describing respectively the reference arm and  
the arm containing the dephasee.
If the dephasor-dephasee interaction is switched off
both terms are product states, 
$\mid \Psi_{ref} \rangle =\mid \phi_{ref} \rangle \otimes \mid \chi_{0} \rangle$ 
and $\mid \Psi_{d} \rangle =\mid \phi_{0} \rangle \otimes \mid \chi_{0} \rangle$,
where $\mid \chi_{0} \rangle$ is the initial state of the dephasor and
$\mid \phi_{ref} \rangle$, $\mid \phi_{0} \rangle$
 are the states in the two arms of the interferometer.
In the presence of the interaction $\mid \Psi_{d} \rangle$ 
is an {\it entangled} state
$\sum_{mn}a_{mn}~\mid \phi_{m} \rangle 
\otimes \mid \chi_{n} \rangle$ where the 
sum, \it far away from the interaction region\rm, 
is restricted to the energetically available states of the subsystems.

The dephasing is related to the reduction from unity 
of the modulus of the elastic amplitude $a_{00}$. 
For the particular case of a "rigid" dephasee,
whose initial state does not change as a result of the interaction, 
$\mid \Psi_d \rangle 
= \mid \phi_{0} \rangle \otimes \sum_{n}a_{0n}\mid \chi_{n} \rangle =
\mid \phi_{0} \rangle \otimes \mid \chi_{fin} \rangle$ and  $a_{00}$ is given 
by a simple
overlap $a_{00}= \langle \chi_{0} \mid \chi_{fin} \rangle$. In this case, 
only a dephasor with 
degenerate initial state can cause dephasing as the ``rigid'' 
constraint forbids any exchange of energy between the two subsystems.
The number of open inelastic channels is then proportional to 
the degeneracy of the initial state of the dephasor.
However, in the general case as a result of entanglement
the number of open inelastic channels depends
on both subsystems and \bf not only \rm
on the dephasor. 
The above simple considerations are clearly  schematic and applicable for an
idealized  pure state situation. They are however helpful to elucidate
also more realistic cases (cf., below).   

In this paper, we will illustrate  the previous discussion
using the WPD experiment as an example.
In the WPD, the flux-sensitive part of the transmission probability
through the interferometer is 
\begin{equation}
{\cal P}_{AB} \propto\; {\rm Re}\; \left[ t_{ref}\;  
t_{QD}\; e^{\;2 \pi\; i \;\bf \Phi/ \Phi_0}\right]~ ,
\label{1.1}
\end{equation}
where $t_{ref}$ is the transmission amplitude~\cite{foot} through the arm
without QD  and 
$t_{QD}$ is the elastic amplitude to cross the arm containing
the QD. Referring to the preceding general discussion  $t_{QD}$ is proportional
to the  amplitude $a_{00}$.   

It is convenient to characterize the dephasing  by means of a positive
function $\nu$, defined as 
\begin{equation}
\mid t_{QD} \mid= \mid t_{QD}^{0} \mid(1-\nu \;(eV,{\cal T}))~.
\label{1.1a}
\end{equation}
where $\mid t_{QD}^{0} \mid$ is the transmission amplitude  when the 
interaction with the QPC is switched off, ${\cal T}$ is the transparency of
the QPC, and $eV$ is the bias applied to it.

The bias, transparency, and the temperature dependence  
of $\nu$ have recently been investigated both experimentally~\cite{Bucks}
and theoretically~\cite{Gurwitz,Yeoshua,Aleiner,Buttiker,Hackembroich}.
The theoretical results indicated that for temperatures $T<eV$
\begin{equation}
\nu \;\simeq \; \;2 \pi \lambda_d^2\frac{\mid eV \mid}{\Gamma}~.
\label{old}
\end{equation}  
Here  $\lambda_d$ is the coupling between QD and QPC. It is conventionally
expressed as~\cite{Bucks,Yeoshua} 
\begin{equation}
\lambda_d^2 \simeq \frac{1}{16 \pi^2}\frac{(\Delta{\cal T})^2}{ {\cal T}(1-{\cal T})}~,
\label{coupling}
\end{equation}
where $\Delta {\cal T}$
is the change in the transparency due to the presence of an
electron in the QD. 
These results were tested experimentally~\cite{Bucks} in the following regime
\par (i) the QD was tuned to the maximum of a Coulomb Blockade (CB) peak and
\par (ii) the temperature was much higher than the coupling to the leads, 
$T \gg \Gamma$.

At such temperatures the number of inelastic channels is mainly determined 
by $T$ and $eV$. In order  to investigate the  dephasor-dephasee interplay
we  focus in this work on the opposite low-temperature limit,  $T \ll \Gamma,eV$.
In Section~\ref{mod} we present the details of our calculation and derive a  
general expression for $\nu$ (Eq.~(\ref{1.15})-~(\ref{1.17})). 
Here we summarize our main findings. 

According to the standard description of a QPC~\cite{Lesovik},
the bias $eV$ represents the difference in chemical potentials between left and right
going scattering states. Therefore, the two energy scales of the problem
$\Gamma$ and $eV$ characterize the windows of states in the QD and QPC 
which can participate in mutual scattering processes leading to dephasing. 
The idealized ``rigid'' dephasee
limit is $\Gamma\rightarrow 0$ for which  
the states in the QPC can only scatter to states with the same energy. 
The number of open inelastic channels is then proportional to $eV$.
Therefore, the calculation of $\nu$ performed in this 
limit~\cite{Yeoshua,Aleiner,rate} gives the 
result of Eq.~(\ref{old}).
For a finite  $\Gamma$ 
we find that the ``rigid'' approximation   
is valid \bf only \rm in the limit of $\Gamma \ll eV$. 

As the bias is reduced, the function $\nu$ crossover smoothly to
the opposite regime ($eV \ll \Gamma$) where the dephasee cannot be approximated
as ``rigid''. The scattering in QPC can  occur between any states in the 
$eV$ interval and the number of open inelastic channels is proportional 
to $(eV)^2$. In this case we find accordingly  
\begin{equation}
\nu\;
\simeq \;2 \lambda_d^2
\left(\frac{eV}{\Gamma}\right)^2 ~.
\label{new}
\end{equation}
Thus in this regime the dephasor-dephasee interplay 
significantly  
modifies the "rigid" dephasee result leading in a sense to observable effects
of entanglement on dephasing. 
In practical terms it also means that 
as a consequence of entanglement the dephasing ability of a 
dephasor is not universal but depends  on the particular properties of the 
dephasee.

In Section ~\ref{mod} 
we study as well the dependence of $\nu$ on the detuning of 
the QD away from the maximum of a CB peak and show that also in this respect 
the two regimes discussed above are significantly distinct (cf., Eq~(\ref{1.15a})
and Eq~(\ref{1.15b})).   

\section{The model and calculation}
\label{mod}

In order to describe the WPD, we can focus  on the arm containing 
the QD.  If  $\Gamma$ and $eV$ are less than the mean level spacing in the QD
we can represent it by a resonant level model
\begin{equation}
\hat{H}_{QD}=\epsilon_0 \;d^{\dagger}_0 d_0+\sum_{k,i}\;
\left(\epsilon_k\;c^{\dagger}_{ki}c_{ki}+ (A_{ki}c^{\dagger}_{ki}d_0+c.c)
\right)
\label{1.1b}
\end{equation}
where the operators $c_{ki}^{\dagger}$ and $c_{ki}$ refer to
states in the left-right leads ($i=L,R$) connected to the QD, and 
the operators $d_0^{\dagger}$, $d_0$ to the resonant level 
in the QD. 

We describe the QPC 
by means of the standard picture in terms of 
scattering states~\cite{Lesovik}. For simplicity we assume that
a single transversal channel is open. Therefore
the total Hamiltonian is
$\hat{H}=\hat{H}_{QD}+\hat{H}_{QPC}+\hat{V}$ with~\cite{Yeoshua,Aleiner}
\begin{eqnarray}
&& \hat{H}_{QPC}=\sum_{q,j}\; \epsilon_q\;b^{\dagger}_{qj}b_{qj}~,\nonumber\\
&& \hat{V}=d^{\dagger}_0 d_0\; \sum_{j,j^{\prime}}
\sum_{q,q^{\prime}}\;V_{qq^{\prime}}
(j,j^{\prime})\;
b^{\dagger}_{qj}b_{q^{\prime}j^{\prime}}~,
\label{1.2}
\end{eqnarray}
where the operators $b^{\dagger}_j$ ($b_j$)
create (annihilate) left-right going scattering states 
($j=L,R$).
The application of a bias shifts the chemical potentials 
of the right-left going scattering states such that $\mu_R-\mu_L=eV$.
Moreover we take
the matrix elements
$V_{qq}(j,j)$ to zero if
$\epsilon_q<\mu_{L}$. This corresponds to the inclusion   
of the equilibrium Hartree shift in the definition of $\epsilon_0$.

We are interested in the modification of $t_{QD}$ due to the presence of
the QPC. This transmission amplitude is 
\begin{equation}
t_{QD}=-i \sqrt{4\Gamma_L\Gamma_R} \int d\omega f^{\prime}(\omega)
\it G_{d} \rm (\omega)~,
\label{1.3} 
\end{equation} 
where $f^{\prime}$ is the derivative of the Fermi function and
$\Gamma_{L (R)}$ is the elastic coupling to the left
(right) reservoirs
($\Gamma_i=\pi \sum_k \mid A_{ki} \mid^2 \delta(\omega -\epsilon_k)$)
assumed energy independent. 

Information about the QD and its interaction with the QPC is contained in 
$\it G_{d} \rm (\omega)$ which is the Fourier transform of the retarded 
Green's function, $\it G_{d} \rm (t)=-i\; \theta(t)\;
\langle \{ d_0(t),d^{\dagger}_0(0) \}\rangle $.
In order to calculate it, we use a real time perturbation expansion 
in the interaction $\hat{V}$ between the QD and the QPC.
The general result is
\begin{equation}
\it G_{d} \rm (\omega)=\frac{1}
{(\omega-\epsilon_0)+i \Gamma - \Sigma (\omega)} ~,
\label{1.6}
\end{equation}
where $\Gamma=\Gamma_L+\Gamma_R$ and $\Sigma$ is the proper 
retarded self-energy.
{\setlength{\unitlength}{1cm}
\begin{figure}
\begin{picture}(4,3.3)
   \put(-1,-7){\psfig{width=10cm,figure=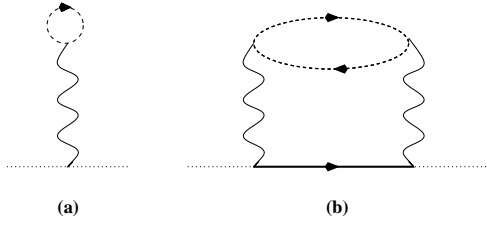}}
\end{picture}
\caption{Lowest order contributions to $\Sigma$. The bold line
represent the propagator in the QD while the dashed lines represent propagators
in the QPC.}
\label{Fig1}
\end{figure}}
In the weak coupling limit,   
we can approximate the self energy with 
the diagrams of Fig.~\ref{Fig1}. 
We expect this approximation to describe correctly the scattering 
processes leading to dephasing, provided that at 
every step of the expansion the broadening 
of the QD resonant level due to the coupling to the leads
is taken into account.
Therefore, we calculate the diagrams of
Fig.~\ref{Fig1} using for the QD the noninteracting Green's functions   
\begin{eqnarray}
&&G_0^<(\omega)=i f(\omega) A_0(\omega) ~,\nonumber \\
&&G_0^>(\omega)=-i (1-f(\omega)) A_0(\omega)~,
\label{1.10a}
\end{eqnarray}
where $A_0(\omega)=2\Gamma/((\omega-\epsilon_0)^2+\Gamma^2)$ 
is the noninteracting spectral density.
The analogous Green's functions of the QPC are
\begin{eqnarray}
&&g^<_{qj}(\omega)=2\pi i f^{(j)}(\omega) \delta(\omega-\epsilon_q)~,
\nonumber \\
&&g^>_{qj}(\omega)=-2\pi i \left(1-f^{(j)}(\omega)\right)
 \delta(\omega-\epsilon_q)~,
\label{1.10b}
\end{eqnarray}
where the thermal factor $f^{(j)}(\omega)=f(\omega-\mu_{j})$
distinguishes left and right-going scattering states by their
chemical potential difference. 

The first order term (Fig.~\ref{Fig1}a)
provides a shift of the QD level 
\begin{equation}
\Delta=2\;\sum_{q,j}\;V_{qq}(j,j)\;f^{(j)}\left(\epsilon_q\right)~,
\label{1.7}
\end{equation}
the factor of two coming from the summation over spin  in the QPC.

An important contribution to $\Sigma$ comes from the second order
bubble diagram (Fig.~\ref{Fig1}b) describing the 
scattering between electrons in the QD and in the QPC. This contribution
can be written as
\begin{equation}
\Sigma_{2}(\omega)=i\int \frac{d\omega^{\prime}}{2\pi}\;
\frac{\Sigma^{>}_{2}(\omega^{\prime})-\Sigma^{<}_{2}(\omega^{\prime})}
{\omega-\omega^{\prime}+i\delta} ~,
\label{1.8}
\end{equation}
where the self energies $\Sigma^{>}$ and $\Sigma^{<}$ 
can be expressed in terms 
of the greater and lesser Green's functions~\cite{Mahan}
of the QD and of the QPC as
\begin{eqnarray}
\Sigma^{<}_{2}(\omega^{\prime})&=& 2\;\sum_{i,j}\sum_{q,q^{\prime}}
\mid V_{qq^{\prime}}(j,j^{\prime})
\mid^2 \int \frac{d \omega_1 d\omega_2}{(2 \pi)^2}
\nonumber \\
&&G_0^<(\omega^{\prime}-\omega_1) g^<_{qj}(\omega_2) g^>_{q^{\prime}j^{\prime}}
(\omega_2-\omega_1)~,
\label{1.9}
\end{eqnarray}
and a similar equation  for $\Sigma^{>}$ with interchanged  symbols $<$ and 
$>$ . 
Substituting the expressions for the Green's functions into Eq.~(\ref{1.9}), 
after some straightforward algebra, we arrive at the result
\begin{eqnarray}\label{sigma}
\Sigma_{2}(\omega)&=&2 \int\;\frac{d\omega'd\omega''}{2 \pi}\rho_0(\omega'')
\Big[\; \frac{A_0(\omega')f(\omega')}{\omega+\omega''-\omega'+i\delta}\\ \nonumber
&& +\frac{A_0(-\omega')f(\omega')}{\omega-\omega''+\omega'+i\delta} \;\Big], 
\end{eqnarray}
where the spectral density
\begin{eqnarray}\label{rho}
\rho_0 (\omega)&=&\sum_{j,j^{\prime}} \; \lambda^2(j,j^{\prime}) 
\int d\omega_1 d\omega_2\; \delta(\omega_1-\omega_2-\omega)\nonumber \\
&&f^{(j)}(\omega_2)\left(1-f^{(j^{\prime})}(\omega_1)\right) ~,
\end{eqnarray}
describes particle-hole excitations in the QPC. 
Here we introduced the dimensionless couplings $\lambda(j,j')=L\mid V(j,j)
\mid/(2\pi v_F)\ll 1$.

We are interested 
in $T \ll \Gamma,eV$  which amounts to taking  
the limit $T \rightarrow 0$. From Eq.~(\ref{1.3}),(\ref{1.6}) it follows that 
in order to calculate the transmission amplitude $t_{QD}$ 
one needs $\Sigma (\omega)$ at the Fermi level $\epsilon_f$ 
in the the arm containing the QD. In the following we will set $\epsilon_f =0$
without loss of generality. 
The real and imaginary parts 
of $\Sigma_2$ at the Fermi level can be written as
\begin{eqnarray}\label{1.12a}
&&\Delta_d\equiv\text{Re}[\Sigma_2]=\;P \int \frac{d\omega' d\omega''}{\pi}
\frac{\rho_0 (\omega'') A^-(\omega')}
{\omega''-\omega'}f(\omega')~,\\
&&\Gamma_d\equiv-\text{Im}[\Sigma_2]=\int d\omega' 
f(\omega') A^+(\omega') \rho_0 (\omega')~,
\label{1.12}
\end{eqnarray} 
where $A^{\pm}(\omega)=A_0(\omega) \pm A_0(-\omega)$.

From  Eq.~(\ref{1.12a}) it follows that, 
because of the symmetry of $A_0(\omega)$,
$\Delta_d$ goes to zero as $\epsilon_0$ goes to zero.
For $\epsilon_0<\Gamma$ we can estimate
\begin{equation}\label{useless}
\Delta_d \simeq -\lambda_e^2 
\epsilon_0 (\xi_0/\Gamma) ~,\nonumber
\end{equation}
where $\lambda_e=\lambda(j,j)$ and $\xi_0$ is a cutoff
of the order of the Fermi energy in the QPC.
Therefore 
$\epsilon_0+\Delta_d \simeq (1-\lambda_e^2 \xi_0/\Gamma) \epsilon_0$
from which we see that in the weak coupling regime this shift 
can be neglected.

The imaginary part $\Gamma_d$ controls the dephasing 
caused by the interaction with the  QPC. 
At zero temperature and with the QPC in equilibrium the spectral density
$\rho_0$ is finite only for positive frequencies. As expected, this gives vanishing 
$\Gamma_d$. 
If a finite bias is applied to the QPC we have 
\begin{equation}
\rho_0 (\omega)=\lambda_d^2\;(\omega+eV)\;\;\;\rm \text{for}\;\;  \omega<0 ~,
\label{1.14}
\end{equation}
where we used the notation $\lambda(j,j^{\prime})=\lambda_d$ for
$j \neq j^{\prime}$.
Substituting Eq.~(\ref{1.14}) in Eq.~(\ref{1.12}) and performing the
frequency integration we obtain  
\begin{eqnarray}
\Gamma_d &=& 
\lambda_d^2\Gamma\;\bigg\{\ln\left(\frac{(1+\xi_0^2)^2}
{(1+\xi_+^2)(1+\xi_{-}^2)}\right)\nonumber\\
&& +2\;\left(\xi_{+}\right)
\left[\arctan\left(\xi_{+}\right)
-\arctan \left(\xi_0\right)\right]\nonumber\\
&& + 2\;\left(\xi_{-}\right)
\left[\arctan\left(\xi_{-}\right)
+\arctan \left(\xi_{0}\right)\right]\bigg\}~,
\label{1.15}
\end{eqnarray}
where we introduced the dimensionless variables $\xi_{\pm}=(eV \pm \epsilon_0)/\Gamma$
and $\xi_0=\epsilon_0/\Gamma$.
The dependence of $\Gamma_d$ on the transparency
${\cal T}$ of the QPC is contained in the coupling constant $\lambda_d$
(Eq.~(\ref{coupling})). 

Using Eq.~(\ref{1.15}) together with Eq.~(\ref{1.3}) and
Eq.~(\ref{1.6}) we finally obtain the 
transmission amplitude
\begin{eqnarray}
t_{QD}=\frac{i \sqrt{\Gamma_L\Gamma_R}}{(\epsilon_0+\Delta)-i(\Gamma
+\Gamma_d)}~,
\label{1.16}
\end{eqnarray}
where $\Delta=2 \lambda_e eV$ ($\lambda_e=\lambda(j,j)$) can be easily 
obtained from Eq.~(\ref{1.7}).
In this formula both $\Delta$ and $\Gamma_d$ renormalize the transmission 
amplitude. However, the effect of $\Delta$ 
is not related to dephasing and in real experiments 
$\epsilon_0$ is typically (e.g., by means of a plunger
gate) so that $\epsilon_0+\Delta_0=\epsilon$ is maintained constant.


It is now possible to relate $\Gamma_d$ to the suppression
of Aharonov-Bohm oscillations $\nu$ defined in Eq.~(\ref{1.1a}).
We write $\mid t_{QD} \mid$ in terms of the renormalized
level $\epsilon$ and expand it up to second order in $\lambda$.
The result is
\begin{eqnarray}
\nu =\frac{\Gamma^2}
{\Gamma^2+\epsilon^2}\;\left(\frac{\Gamma_d}{\Gamma} \right)~.
\label{1.17}
\end{eqnarray}

When the amplitude of A-B oscillations is measured, the relevant scale
on which $\epsilon$ is varied near the Fermi level is
$\Gamma$. In this interval 
one can obtain from Eq.~(\ref{1.15}) two simple expressions
for $\Gamma_d$ in the opposite limits of large and small bias
\begin{eqnarray}\label{1.15a}
\Gamma_d \;&\simeq& \;2 \lambda_d^2
\;\left[\frac{\Gamma^2}{\Gamma^2+\epsilon^2}\right]\;
\frac{(eV)^2}{\Gamma} \;\;\;\;\;\text{for}\;\;\; eV \ll \Gamma~,\\
\Gamma_d \;&\simeq& \;2 \pi \lambda_d^2 \mid eV \mid 
\;\;\;\;\;\;\;\;\;\;\;\;\;\;\;\;\;\;\;\;\;\text{for}\;\;\; eV \gg \Gamma~. 
\label{1.15b}
\end{eqnarray}
Substituting these expressions in 
Eq.~(\ref{1.17}) with $\epsilon=0$ we recover the two results
anticipated in Section \ref{intr} (Eq.~(\ref{old}) and Eq.~(\ref{new})).
One moreover observes that the QD-QPC entanglement leaves another fingerprint in 
the dependence on the detuning of the QD resonance $\epsilon$.
Indeed, for $eV \ll \Gamma$ the function
$\Gamma_d$ is very sensitive to the variations of $\epsilon$ 
, while in the other case these variations are negligible.  

In Eq.~(\ref{1.16}) the energy scale $\Gamma_d$ appears as a correction to the 
width $\Gamma$. However, we emphasize that $\Gamma_d$ is the imaginary part of 
$\Sigma_2$ at a specific energy ($\omega=0$). Examining the energy dependence
of $\text{Im}[\Sigma_2]$ we discover another profound difference 
between the two regimes. Taking for simplicity $\epsilon_0=0$ we obtain  
$\mid \text{Im}[\Sigma_2(\omega)] \mid \approx \Gamma_d + 
2 \lambda_d^2 \omega^2/\Gamma$ for $\mid \omega \mid \lesssim \Gamma$.
In this interval the relative variation of $\text{Im}[\Sigma_2]$, 
defined as $\delta \equiv \mid \text{Im}[\Sigma_2(\Gamma)]+\Gamma_d \mid/ \Gamma_d$,
is given by
\begin{eqnarray}  
\delta &\approx& \left(\frac{\Gamma}{eV}\right)^2 
\gg 1\;\;\;\;\;\;\text{for}\;\;eV \ll \Gamma\\
\delta &\approx& 
\frac{\Gamma}{\mid eV \mid} \ll 1\;\;\;\;\;\;\;\;\;\text{for}\;\;eV \gg \Gamma
\label{1.19}
\end{eqnarray}
It therefore follows that only 
in the large bias case (``rigid'' dephasee limit) $\text{Im}[\Sigma_2]$
is well approximated by a constant $\Gamma_d$. 
Only then it is possible to interpret $\Gamma_d$ as a \it dephasing rate \rm 
and the dephasing-induced time dependence of the QD Green's
function is exponential ($G_d(t)\propto \exp[-(\Gamma+\Gamma_d)\;t]$). 
In the opposite regime the strong variations 
of $\text{Im}[\Sigma_2]$ with $\omega$
imply that the characterization of the dephasing
by \it dephasing rate \rm\,~\cite{Yeoshua,Aleiner,Lesovik2}
is \it impossible \rm, i.e., the dephasing is 
\it not \rm described by a simple exponential decay of $G_d(t)$.

We would like to express special thanks to Y. Levinson for many useful 
and stimulating discussions.
We are also thankful to Y. Imry, 
M. Heiblum, B. Rosenow, D. Sprinzak, I. Ussishkin, M. Rokni, 
and M. Schechter for valuable discussions. We acknowledge the support
of the DIP grant no.DIP-C7.1.

\widetext

\begin{references}
\bibitem{rate} This is done in Ref.\ \onlinecite{Yeoshua}
, and Ref.\ \onlinecite{Aleiner} using the limit $\Gamma \rightarrow 0$
for the calculation of the dephasing rate $\Gamma_d$.
An expression for $\nu$ can then be obtained 
through the formula $\nu =\Gamma_d/\Gamma$. 
(see Eq.~(\ref{1.17}) and the discussion following).
\bibitem{foot} We assume here that $t_{ref}$ is energy independent.
\end{references}
\end{document}